# Achromatic single-layer hologram

**Zhi Li[1,2], Wenhui Zhou[1], Xin Yuan[3], Weiwei Cai[4], Dongdong Teng[2], Qiang Song[1*], and Huigao Duan[1*]**

[1]Greater Bay Area Institute for Innovation, Hunan University, Guangzhou 511300, China
[2]School of Physics, Sun Yat-sen University, Guangzhou 510275, China
[3]School of Engineering, Westlake University, Hangzhou 310030, China
[4]Key Lab of Education Ministry for Power Machinery and Engineering, School of Mechanical Engineering, Shanghai Jiao Tong University, Shanghai 200240, China
*Corresponding author: songqiangshanghai@foxmail.com; duanhg@hnu.edu.cn

**Abstract**. Phase retrieval is a fundamental technique of advanced optical technologies, enabling precise control over wavefront properties. A persistent challenge in diffractive optical element (DOE) design is that a single hologram typically operates within a single wavelength or color channel, limiting it to monochromatic image generation. This limitation in channel capacity significantly restricts the applicability of DOE in optical applications. In this study, we propose a design strategy for full-color, single-layer hologram based on a variable-scale diffraction model. By imposing strict constraints in Fourier domain and reducing depth of focus (DOF), we achieve the simultaneous encryption and storage of red, green, and blue channel information within a single achromatic hologram. This strategy facilitates color separation in large-depth 3D holography and enables achromatic full-color image displays. We demonstrated full-color holographic video playback at a full refresh rate of 60 Hz, achieving a temporal resolution three times greater than that of existing methods. Furthermore, we successfully fabricated achromatic, twin-image-free, full-color binary pure-phase DOEs at low cost. This achromatic strategy addresses the demands across various fields in optics, including high-refresh-rate full-color displays, high-density optical information storage, advanced optical security, high-reusability holographic metasurface optical element, and high-performance achromatic metalenses.

**Keywords**: phase retrieval, achromatic hologram, information capacity, diffractive optical element.

## Introduction

Wavefront control is a crucial technology in optics community. By precisely modulating properties such as the phase and amplitude, it optimizes the propagation characteristics of light beams. This capability has significantly advanced high-resolution imaging and display technologies [1–6] while also playing a key role in applications like optical security[7–9], sensing[10,11], and the design of micro optical devices[12–15]. Holography, a representative wavefront modulation technique, encodes information from complex images or objects into a thin film known as a hologram. Traditional optical-recording holograms require physical objects for their creation, but they are expensive to produce and offer limited viewing angles[8,9]. In contrast, computer-generated holography (CGH), with its controllable diffraction distances, eliminates the need for physical objects. It is a phase-retrieval technique under bilateral constraints that can achieve desired field distributions at specific locations through wavefront modulation. By integrating neural network algorithms and optimization techniques, CGH is widely applied in areas such as high-quality holographic displays[16–18], polarization-multiplexed metasurfaces[12,19,20], beam shaping[21,22], and achromatic metalens design[23–25].

Despite significant progress in phase retrieval algorithms, particularly in display quality, generation speed, and high-precision device manufacturing, a persistent challenge remains. A single hologram generated by phase retrieval algorithms, or scalar diffraction field calculated based on a single object, is typically customized for only one wavelength. This limitation arises due to the coherent interference requirements of diffractive optics, which restricts wavefront modulation to a single channel. In color holography, the red, green, and blue channels are often switched to achieve a full-color display. For devices, multiple polarization-insensitive diffractive optical elements (DOEs) or polarization-multiplexing metasurfaces which contain three or more holograms are required to achieve full-color imaging. However, these methods increase system complexity and cost, making single-layer solutions highly desirable. A DOE with pure-phase characteristics serves as a typical two-dimensional rigid carrier for static CGH and is widely used in imaging[26,27], automotive components[28], beam-shaping devices[29,30], and optical amplifiers[31]. Compared to holographic metasurfaces, pure-phase DOEs offer several advantages: simpler structures, lower production costs, greater physical stability, and easier characterization. For instance, polarization-insensitive DOEs used in anti-counterfeiting applications enhance security, simplify integration, and enable the display of encrypted information while significantly reducing mass production costs. Imitating these DOEs without expertise in lithography fabrication systems and design algorithms is nearly impossible, making them highly valuable in optical security applications due to their complexity. However, a single-layer pure-phase DOE is typically limited to carrying only one hologram, constrained by its single-channel design. This is one of its primary disadvantages compared to polarization-multiplexed metasurfaces, which can manage multiple channels simultaneously.

If a single hologram could carry the full channel capacity, that is, store and manipulate the information for red, green, and blue channels simultaneously, it would dramatically reduce system complexity and cost. Such an approach would eliminate the need for multiple layered elements, simplify fabrication processes, and improve integration in optical systems. Moreover, it would enable the development of high-efficiency, full-color holographic displays and devices. This breakthrough could also expand the phase retrieval algorithms in applications like optical security, data storage, and real-time displays, where color information capacity is crucial.

To achieve this function, we propose a design pipeline for single-layer full-color holograms. This method reduces chromatic aberration and information crosstalk by employing variable-scale diffraction algorithms, image plane constraints, and techniques that reduce depth-of-focus (DOF) issues. Notably, our approach does not require image preprocessing or the use of blazed grating phases[32]. A single hologram can simultaneously modulate red, green, and blue light to generate accurate color information at specific depths. This approach offers high flexibility, superior fidelity, and minimal crosstalk, making it particularly suitable for applications in security and data storage. Under this pipeline, large-scale 3D color-separation holography and full-color holography using a single hologram have been successfully demonstrated. Additionally, we have achieved a full-color 60Hz holographic video with three times the channel capacity of existing algorithms[33]. Moreover, using a two-level lithography process, we fabricated twin-image-free, achromatic, single-layer security DOEs. These polarization-insensitive DOEs can generate full-color images at specified diffraction distances under vertically white light illumination, which is mixed by red, green, and blue lasers, as illustrated in Fig. 1. This method represents a significant advancement in achromatic wavefront control, providing new solutions for full-color holography, optical security, and information storage, which are areas that have long sought enhanced color performance and increased information capacity.

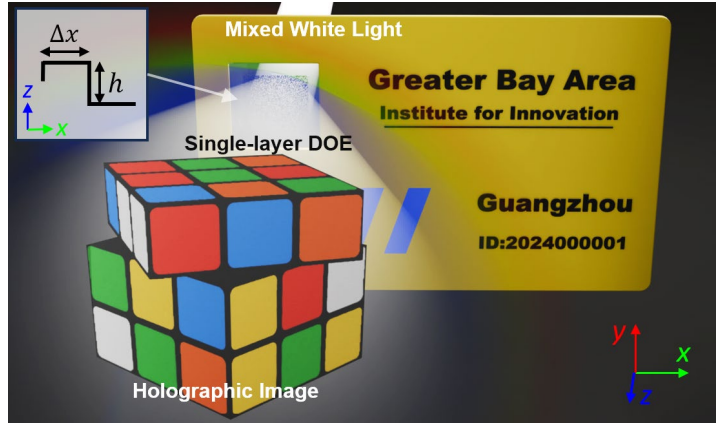

**Fig. 1** Schematic of full-color imaging using achromatic single-layer security DOE. The DOE is embedded in an ID card and illuminated by mixed white light. After modulation by the pixels on the DOE, the light generates a full-color image at a designated diffraction distance. The inset illustrates the two-level section, where $\Delta x$ denotes the pixel size and $h$ represents the step height.

## Methods

The challenges for achromatic single-layer hologram design focus on three key aspects. i) The algorithms must ensure high-fidelity identification information at specific depths and positions while minimizing chromatic aberration and the twin-image effect. ii) The size of the image at the targeted depth must be precisely controllable to meet various identification or display requirements. iii) It is crucial to minimize unwanted information crosstalk from images on other diffractive planes while maintaining the integrity of identification data, thereby preventing information leakage or misinterpretation.

The phase retrieval algorithms depend on fast computational methods for scalar diffraction[34]. The widely used angular spectrum method necessitates that the image plane size match that of the hologram and is limited to near-field applications. Additionally, the single Fresnel transform, which exhibits magnification proportional to the wavelength $\lambda$ for far-field propagation, can introduce chromatic aberration and typically requires preprocessing that compromises image fidelity. In contrast, the spatial frequency tunable method (SFTM)[35], derived from phase space analysis[36,37], allows for arbitrary adjustment of the spatial frequency of the image plane. This enables the calculation of diffraction images at any diffraction distance $z$ and scale, subject to the constraints of the sampling theorem. The SFTM is obtained by decomposing the propagation matrix $\boldsymbol{T} = \begin{bmatrix} 1 & \lambda z \\ 0 & 1 \end{bmatrix}$ of the Wigner distribution function in phase space:

$$\boldsymbol{T}_{\text{SFTM}} = \boldsymbol{Q}\left[\frac{m-1}{m\lambda z}\right] \boldsymbol{M}[m] \boldsymbol{F}^{-\pi/2}[1] \boldsymbol{Q}\left[-\frac{\lambda z}{m}\right] \boldsymbol{F}^{\pi/2}[1] \boldsymbol{Q}\left[\frac{1-m}{\lambda z}\right], \qquad (1)$$

where $m$ is the scaling factor of the image plane size relative to the hologram plane size, and $\boldsymbol{M}[m]$ represents the scaling matrix. $\boldsymbol{Q}[c]$ corresponds to the quadratic phase modulation, where $c$ is a non-zero constant. $\boldsymbol{F}^{\pi/2}[c]$ and $\boldsymbol{F}^{-\pi/2}[c]$ correspond to the Fourier transform and inverse Fourier transform, respectively. In a single-pass computation using SFTM, as long as each step adheres to the sampling theorem, adjusting the scaling factor $m$ allows for generating a diffraction image that is $m$ times the size of the input plane at any desired diffraction distance. For a given wavelength $\lambda$, the combination of $m$ and $z$ values that meet all sampling requirements is defined as the *m-z* space. A more detailed explanation of SFTM is provided in Supplementary Material. The design of achromatic single-layer holograms is based on SFTM, since SFTM possesses the ability to automatically correct chromatic aberration.

The SFTM ensures chromatic aberration correction at the fundamental level of diffractive calculations, thereby maintaining image fidelity. However, in the case of three-channel applications, such as full-color image encryption, the hologram generated by superimposing the red, green, and blue holographic sequences results in 9 (3×3) image components, leading to crosstalk among the images. We propose that the achromatic phase retrieval algorithm should impose strong constraints on the image plane in the vertical axis, or Fourier domain, while reducing constraints in the longitudinal (*z*) direction, reducing the DOFs. In bilateral constrained phase retrieval problems, amplitude constraints are generally applied on the image plane. In our approach, the constraints on the image plane are set as a diffuse constraint with scattering properties:

$$\boldsymbol{u}_z^{\text{dif}}(x,y) = \bar{u}_z(x,y) \exp[i\phi_n^{\text{dif}}(x,y)], \tag{2}$$

where $\bar{u}_z(x,y)$ represents the amplitude distribution of the target image, while $\phi_n^{\text{dif}}(x,y)$ is an initial guess for scattering. Physically, scattering surface disrupts the free propagation of light, altering the beam's transmission characteristics. The phase fluctuation range for random scattering is between −*n*π and *n*π. This configuration reduces DOF, while preventing excessive distortion of the amplitude or intensity distribution during hologram superposition. This approach is conceptually similar to using orthogonality of high-dimensional random vectors in the target plane[38], which utilizes the convolution of random phases with a chirp function to mitigate crosstalk. However, that method relies on the central limit theorem and the law of large numbers under a single-wavelength channel, necessitating a nearly infinite number of diffraction planes and maximizing the sacrifice of image quality.

We aim for the image plane to exhibit relatively weak scattering properties, as strong scattering can introduce significant noise in SW during hologram superposition. We find that setting the value of *n* in the initial guess to a range of 0 to 0.2 provides a favorable balance between reducing crosstalk and maintaining image quality in color applications. The value of *n* is empirical and influenced by various factors, such as the size of the image plane. In the initial iteration cycle, for each color channel, a preliminary phase-only hologram is generated using the modified iterative Fourier transform algorithm (IFTA) based on SFTM. In interaction $k_1$, the three holograms are then superimposed to create a single hologram (for simplification, the coordinate labels have been omitted):

$$\boldsymbol{H}_{k_1+1} = \hbar\left\{\sum_{j=r,g,b} \boldsymbol{H}_{j,k_1}^{\text{SFTM}}\left(\boldsymbol{H}_{k_1}, \boldsymbol{u}_{z_j,j}^{\text{dif}}, \lambda_j, z_j, m_j\right)\right\}, \forall (x,y), \tag{3}$$

where *j*=*r*, *g*, *b* represents red, green, and blue colors respectively. $\boldsymbol{H}_{j,k_1}^{SFTM}$ signifies the hologram obtained using the SFTM based modified IFTA algorithm, and $\hbar\{\cdot\} = \exp(i\mathcal{P}\{\cdot\})$ represents a phase transformation, with $\mathcal{P}\{\cdot\}$ indicating the extraction of the phase distribution. After $ks_1$ interactions and reaching a stagnation point, we extract $\boldsymbol{H}_1^S = \boldsymbol{H}_{ks_1+1}$ as the preliminary hologram. This simple superimposition, however, results in significant background noise and chromatic aberration, requiring further optimization.

Adaptive Fourier domain constraints combined with soft quantization offer significant advantages in the generation of multi-level monochromatic holograms[39]. These constraints, derived through a variational method, guide the distribution on the diffraction plane. We aim to refine this approach to facilitate the design of achromatic single-layer holograms. At a specific diffraction distance, the region where the desired image appears is defined as the signal window (SW), while the area outside this region is considered the noise window (NW). In our strategy, we designate a larger area surrounding the target image as the SW to achieve a more stable distribution for all three color channels. Specifically, we treat a rectangular region at the center of the image plane as the SW rather than limiting it to the pixels occupied by the image itself. The target image is then placed within this rectangle, with all pixels in the SW subject to constraints. Within the SW, the constrained distribution $\hat{u}_{z_j,j,k_2}$ on the image plane is modified to:

$$\hat{u}_{z_j,j,k_2} = \begin{cases} K_{j,k_2}\left(K_{j,k_2}\left|\dfrac{u_{z_j,j}^{\text{dif}}}{u_{z_j,j,k_2}}\right|\right)^{\tau} u_{z_j,j}^{\text{dif}}\exp(i\varphi_{j,k_2}), \forall (x,y) \in \text{SW} \\ u_{z_j,j,k}, \forall (x,y) \notin \text{SW} \end{cases}, \quad (4)$$

where $\varphi_{j,k_2}$ is the phase distribution in the image plane in interaction $k_2$, and $K_{j,k_2} = \left(\dfrac{\iint_{SW} |u_{z_j,j,k_2}|^2 dxdy}{\iint_{SW} |\bar{u}_{z_j,j}|^2 dxdy}\right)^{\frac{1}{2}}$ is defined as energy scaling parameter. $\tau$ is a relaxation parameter, which is set to 0.6 in our strategy. In interaction $k_2$,

$$H_{k_2+1}^S = \hbar\left\{\sum_{j=r,g,b} S_j\{H_{k_2}^S\}\right\}, \forall (x,y), \quad (5)$$

where $S_j\{\cdot\}$ denotes the optimization and eight-level soft quantization process under $\hat{u}_{z_j,j,k}$ constraints based on SFTM. After proper $ks_2$ interactions, ultimately, a single hologram is obtained:

$$H = H_{ks_2+1}^S. \quad (6)$$

**Results**

**Channel separation in 3D holography**

A spatial light modulator (SLM) with a pixel size of 4.5 μm is used to demonstrate full-color CGH with phase-only holograms. The holograms have a resolution of 1000×1000 pixels, while the resolution of the SW is 600×600. The area outside the SW constitutes the NW. To verify the color separation capability in 3D holography applications, we project the information carried by each of the three colors at different spatial positions and with varying magnification factors. The designed illumination wavelengths are $\lambda_r$=638 nm, $\lambda_g$=532 nm and $\lambda_b$=450 nm. We project a magnified letter "B" by 1.6 times at $z$=10 cm, a magnified letter "G" by 2.4 times at $z$=15 cm, and a magnified letter "R" by 3.0 times at $z$=22 cm, each at different spatial positions, as shown in Fig. 2(a). The magnification factors lie within the allowable $m$-$z$ space of SFTM. Due to the conservation of radiant flux, information at each color's target position may be "contaminated" by rays from other colors, potentially causing erroneous information or making the expected image indistinct. In simulations, as seen in Fig. 2(b), (c), and (d), the three letters experience a predictable level of interference, but no significant information crosstalk between colors occurs, and the images remain clearly distinguishable. In the experiments, the CGH images are projected onto an optical screen and captured by a CCD camera, rather than being directly imaged by the CCD. This method provides a visual effect closer to human-eye perception in the real world. The experimental results illustrated in Fig. 2(e), (f), and (g) closely resemble the simulated outcomes. Compared to ref. [40], our strategy for 3D holography is not limited to monochromatic images and can be expanded to dimensions of decimeters or even meters. This allows for the possibility of large-area, full-color 3D imaging on a macroscopic scale. Notably, under monochromatic illumination, the image in the corresponding color channel at the designated position remains "uncontaminated," demonstrating a channel-switching function similar to that described in Ref.[41], where different security patterns are displayed under various monochromatic illuminations. However, our method achieves this without the need for complex amplitude modulation.

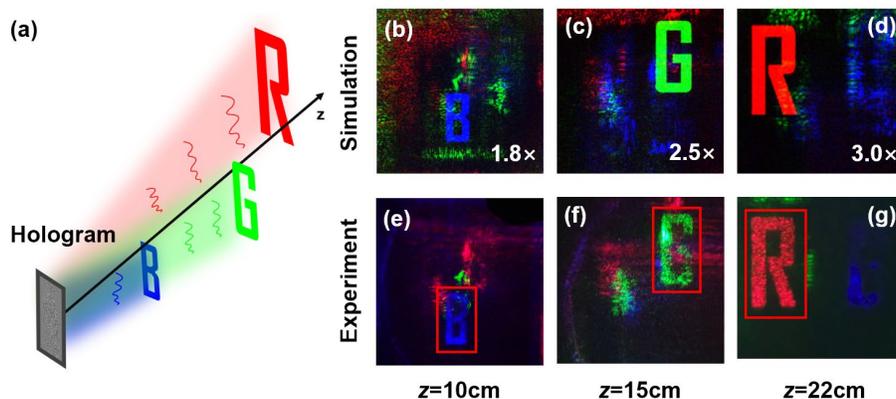

**Fig. 2** Channel separation using a single security hologram. **a** Demonstration diagram of information separation of the three color channels. The hologram is illuminated by mixed white light with $\lambda_r$=638 nm, $\lambda_g$=532 nm and $\lambda_b$=450 nm. **b c d** Simulation and **e f g** experimental results of the three letters "R", "G", and "B" in 3D holography. A magnified "R" by 1.6 times, a magnified "G" by 2.4 times, and a magnified "R" by 3.0 times are projected at different spatial positions at distances of $z$=10 cm, 15 cm, and 22 cm, respectively.

## Achromatic full-color single-layer hologram

The distinguishability demonstrated in Fig. 2 is adequate for basic security applications, as it ensures the accurate display of specific colors and sizes at designated depths, preventing any misleading information under proper verification conditions. However, enhancing the distinctiveness of diffractive patterns by designing full-color holographic images can significantly raise the difficulty of replication, improve recognition efficiency, and offer more unique identification information, thus supporting product premiumization. Moreover, in terms of anti-counterfeiting, our approach does not incur increased material or manufacturing costs like fluorescent materials or films do[42–45].

As illustrated in Fig. 3(a), high-quality full-color CGH is typically realized by temporally switching the SLM to display phase maps for different color components, synchronized with the illumination system. This method leverages a high refresh rate to generate a color image via the persistence of vision. While this approach significantly mitigates chromatic aberration, the gating process significantly increases system costs, reduces the display frame rate to one-third, and crucially, makes it impossible to integrate all three color components into a single, thin hologram layer. An alternative method, shown in Fig. 3(b), is to create a partitioned hologram, where the hologram is divided into three sections, each illuminated by a different laser. However, this approach drastically lowers the image resolution and limits the usable image area[46]. Additionally, aligning the illumination system presents challenges. In recent years, some advances in complex amplitude modulation have emerged, such as combining color filters with holograms to generate different color images under various lasers or laser combinations[41], as shown in Fig. 3(c). However, the nanostructure designs required are highly complex, demanding extreme manufacturing precision, making them unsuitable for common CGH. Another approach utilizes the phase response of cross-shaped meta-atoms and Pancharatnam-Berry phases to decouple two color channels[47]. Yet, this method requires precise polarization control, suffers from substantial crosstalk, and exhibits low diffraction efficiencies.

In our strategy, we place the SWs for the three colors at the same depth and position, and place the red, green and blue components of a color target image in their respective color channels. The magnification factors for the three components are kept consistent to ensure that an accurate full-color image is formed in the end. The use of SFTM based algorithms ensures that the entire full-color image has sufficient freedom in magnification, and the resolution of the three components would not be reduced, unlike with the use of the single Fresnel transform. An enlarged 2.0× cube was projected at a position of $z$=10 cm, with the simulation and experimental results displayed in Fig. 3(e) and (f), respectively. The cube was scrambled, resulting in a more complex color pattern. The simulation reconstruction accurately reproduced the original colors of the cube, and the experimental result was consistent with the simulation. The slight color deviation in the experimental result is partly due to the spatial light modulator's inability to provide completely accurate phase delays for all three colors simultaneously, leading to the actual wavefront that do not perfectly match the design. Projecting white objects with certain details has always been a challenge in holography, as aligning a large range of white pixels across three color channels is very difficult. We attempted to project a complex white object based on our strategy. A 1.8× white cat is imaged at $z$=12 cm, as shown in Fig. 3(g) and 3(h). Both the simulation and experimental results of the cat are clearly distinguishable, and it is nearly impossible to forge a single hologram based on the holographic image in Fig. 3(d).

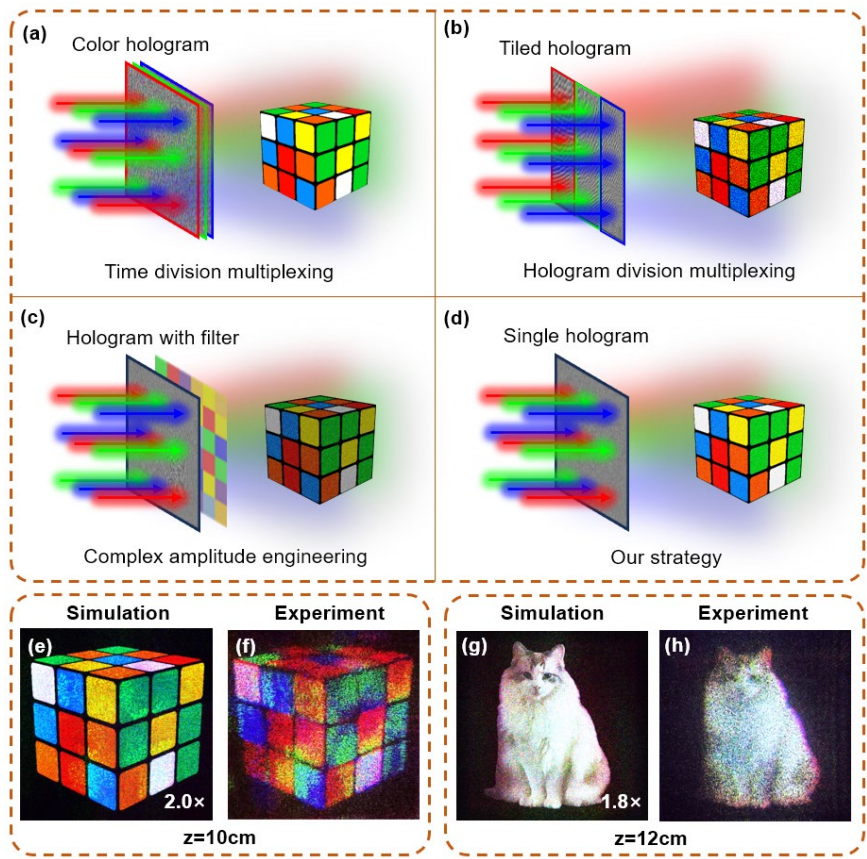

**Fig. 3** Comparison of methods for generating color holographic images and simulation/experimental results of our strategy. **a** Time division multiplexing method. The lasers and their corresponding holograms are switched synchronously, preventing crosstalk between the three color channels. **b** Hologram division multiplexing method. A single hologram is composed of three independent regions, each illuminated by a different laser. **c** Complex amplitude modulation with color filters. High-precision nanostructures are used to simultaneously modulate amplitude and phase to achieve color holographic images. **d** Our strategy. Under mixed white light illumination, a full-color image is obtained using only a single-layer hologram. **e f** simulation and **g h** experimental results. A cube located at $z$=10cm with $m$=2.0 is shown, and a 1.8× magnified white cat image is projected at $z$=12 cm.

**Full-color holographic video with full refresh rate**

Temporal resolution is one of the determining factors for channel capacity. Currently, full-color holographic videos are achieved using a time division method, consistent with the approach shown in Fig. 3(a). As a result, the temporal resolution for color videos is reduced to about one-third of the theoretical maximum, and the effective refresh rate is just one-third of the maximum refresh rate supported by the SLM. For an SLM with a 60 Hz refresh rate, the equivalent refresh rate for full-color holographic videos is only 20 Hz (below the 24 Hz standard for video playback). Although it is possible to achieve 60 Hz full-color holographic video using three SLMs with identical pixel sizes that support 60 Hz, this solution significantly increases the system's cost, size, and alignment complexity. More critically, even with this setup, the channel capacity remains just one-third of the potential.

Our strategy enables the display of a full-color frame using just one phase pattern, thereby tripling the theoretical channel capacity compared to mainstream methods, without the need of electronic shutter devices. We designed and recorded a full-color holographic video (see *Video 1* for simulation payback), demonstrating at $z$=15 cm with a magnification of $m$=2.2. The still images in Fig. 4 show three frames extracted from the simulated video, all of which are full-color images without any segmentation of the red, green, and blue channels. The video was also projected onto a screen and recorded using a mobile device (see *Video 2*). The video refresh rate matches the maximum refresh rate of the SLM, which is 60 Hz, ensuring full refresh rate display and achieving the highest temporal resolution. To the best of our knowledge, this is the first realization of a full-color holographic video with full refresh rate display, providing an efficient solution for high channel capacity, high refresh rate holographic videos and real-time reconstruction.

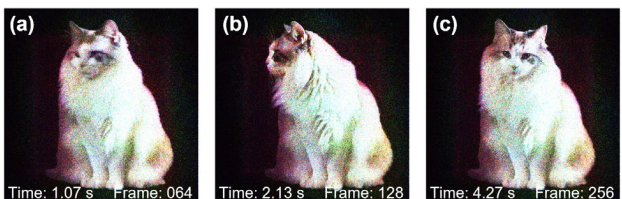

**Fig. 4** Three frames captured from the simulated 60 Hz holographic video, where all the three channels are displayed simultaneously, preserving the full information capacity. **a** Frame 64. **b** Frame 128. **c** Frame 256.

**Achromatic single-layer binary security DOE**

Our strategy offers greater flexibility in the design and fabrication of DOEs, enabling the realization of low-crosstalk, high-feature-definition full-color holography with one single DOE. It opens up new possibilities for the design of polarization-insensitive anti-counterfeiting devices. We aim to apply our strategy to the design of such a polarization-insensitive device. The most cost-effective two-level fabrication process in standard lithography is used to produce the DOE. Although this process may compromise image quality and reduce diffraction efficiency, if the hologram generation algorithm provides a sufficiently high level of anti-counterfeiting security, it can reduce the costs of large-scale manufacturing while still maintaining acceptable image quality.

The base wavelength is $\lambda_r$=638 nm. According to the waveguide effect, for a pixel on the DOE with a height $h$, the phase delay $\phi_j$ provided for $\lambda_j$ is given by

$$\phi_j = (n_j - 1)\frac{2\pi}{\lambda_j}h, \tag{7}$$

where $n_j$ is the refractive index relative to $\lambda_j$. Since the pixels of the DOE cannot provide approximately completely the same phase delay for the three wavelengths, dispersion occurs for different wavelengths, leading to chromatic aberration. To help the phase delays for each color in the DOE align closely with our requirements, the following adjustments for the DOE hologram $H^{\text{DOE}}$ shall be made:

$$\boldsymbol{H}^{\text{DOE}} = \hbar\left\{\sum_{j=r,g,b}\mathcal{A}\left\{\boldsymbol{S}_j\{\boldsymbol{H}^S_{ks_2}\}\right\}\exp\left[i\frac{(n_r-1)}{\lambda_r}\mathcal{P}\left\{\boldsymbol{S}_j\{\boldsymbol{H}^S_{ks_2}\}\right\}\frac{\lambda_j}{(n_j-1)}\right]\right\}, \forall(x,y), \tag{8}$$

where $\mathcal{A}\{\cdot\}$ denotes taking amplitude distribution.

When designing binary DOEs, researchers and engineers typically begin with an off-axis configuration to mitigate the interference of the twin image problem, as the twin image always tends to appear. However, off-axis imaging can still reduce visual quality, as the goal is to present the upright image clearly and accurately at a specific location, while the twin image is undesirable. Additionally, off-axis settings decrease bandwidth utilization and significantly reduces resolution. Moreover, the presence of the twin image can hinder information recognition and the determination of sample authenticity.

In our design of full-color binary DOEs, we discovered an interesting phenomenon that it is possible to visually eliminate the interference of the twin image. Thus, in-line design is performed. For some discussions on twin images refer to the Discussion Section. The DOE has a pixel size of 2 μm × 2 μm, with a total of 5000 × 5000 pixels, resulting in an effective area of 10 mm × 10 mm. This configuration provides sufficient resolution while occupying a reasonable volume on suitable substrates, such as a bank card or automotive holographic projection components.

The DOEs are fabricated using K9 substrates with SPR955-CM photoresist. The thickness of the photoresist on the PDOE is approximately 700 nm. The refractive index measured with an SE-VE-L spectroscopic ellipsometer for red ($\lambda_r$=638nm), green ($\lambda_g$=532nm), and blue ($\lambda_b$=450nm) light are 1.618, 1.634, and 1.655, respectively. Fig. 5 shows the manufacturing process of the DOE samples. The photoresist was spin-coated onto a 1.1 mm thick K9 glass substrate at a speed of 3500 rpm for 60 seconds, which ensures uniform coating and desired thickness. After spin-coating, the sample was heated on a hot plate at 100°C for 90 seconds and then cooled at room temperature for 30 seconds to stabilize the photoresist and reduce internal stresses. The DOE layout patterns were then transferred onto the photoresist using a laser direct writing lithography system (type MiScan-200UV form SVG Tech Group Co., Ltd) with a pulse frequency of 95 kHz. Finally, the sample was immersed in Sun-238 solution for 10 seconds for developing. The sample was immersed in deionized water to remove developer residue from the development process. Then, the sample was placed on the hot plate and baked at 100°C for 3 minutes to complete the hard baking process. Hard baking enhances the adhesion of the photoresist, cures the photoresist, and removes any remaining solvents and moisture, allowing the photoresist to maintain better pattern fidelity for a longer period.

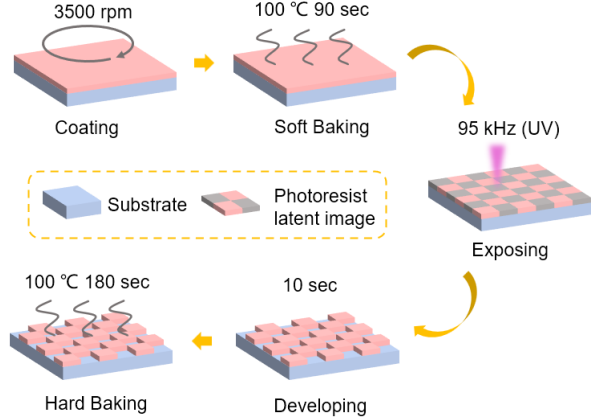

**Fig. 5** Fabrication process of binary DOEs. The entire process is divided into five steps: spin-coating the photoresist, soft baking, photolithography, developing and hard baking. Finally, a protective coating such as polymethyl methacrylate (PMMA) can be applied to extend the sample's lifespan. Notably, the entire process does not require any etching operations.

Fig. 6(a) demonstrates the optical characterization path of the single-layer DOE. A collimated white light, generated by mixing red, green, and blue lasers, vertically illuminates the PDOE embedded in an ID card. The $4f$ system filters out the DC components, and then the image is formed on a screen. Fig. 6(b) shows a top view of the ID card, where the DOE is embedded in the upper-left corner. An optical microscope image of the two-level distribution is presented in the inset of Fig. 6(b). Fig. 6(c) displays the experimental results of the holographic image at $z=10$ cm, revealing a distinct outline of a 2.0× colorful Rubik's cube at the center of the screen, visually free from twin image interference. Additionally, another DOE is designed to project an image of a white cat, located at $z=12$ cm, with a magnification of 1.8×, as shown in Fig. 6(d). Compared to Fig. 3(f), its resolution is higher, and the influence of laser speckle is diminished.

Our full-color DOEs achieve a diffraction efficiency of approximately 30%, which is slightly lower than the theoretical maximum of 40.5% for binary monochromatic DOEs under central symmetry. The absence of twin image allows our DOEs to exceed a diffraction efficiency of 20.25% under in-line settings.

Despite being produced with a two-level process, they are capable of projecting full-color holographic images with high recognizability. Additionally, the hard-baked binary DOE is highly compatible with cover materials such PMMA, preventing scratches and enabling long-term preservation, thereby enhancing their robustness. The high demands of algorithm design, image features, and manufacturing equipment render creating imitations nearly impossible. The low cost of mass production enables this technology to be widely applied in anti-counterfeiting solutions for high-end ID cards, various projection components, and other beam-shaping scenarios.

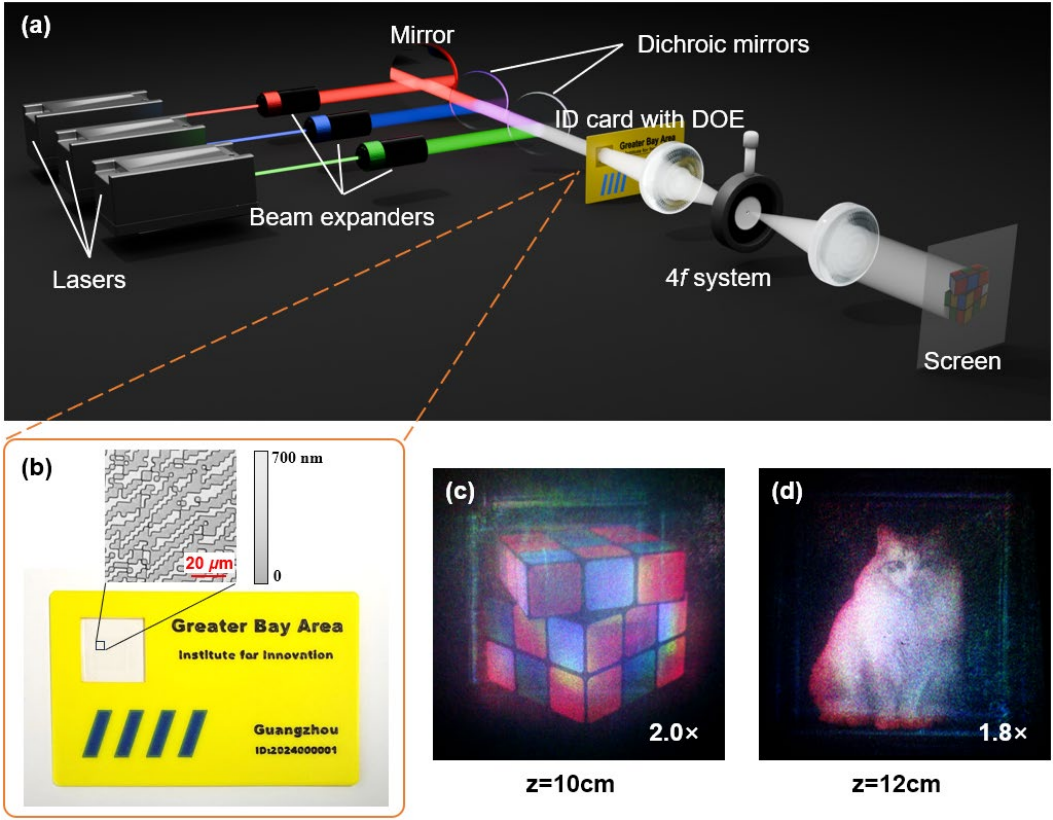

**Fig. 6** Characterization of the DOEs. **a** Experimental setup. The mixed white light vertically illuminates the DOE, and after filtering through a 4*f* system, a full-color image is displayed on the screen. **b** ID card embedded with a single-layer DOE. The inset shows the distribution of two-level structure under an optical microscope. **c d** Full-color holographic images. A colorful cube is displayed at *z*=10 cm with a magnification of 2.0×. In another sample, a white cat is shown at *z*=12 cm with *m*=1.8. Both the two images visually show no significant interference.

## Discussion

The SFTM allows the diffraction plane size to exceed $L_0 + \lambda z / \Delta x$, where $L_0$ is the support size of the input, as the SW area does not exceed this limit and is not affected by aliasing. The size of the SW is sufficiently flexible, and the main concern is the impact of information crosstalk, or DOFs. In our method, while the SFTM ensures the fidelity of the three color channels, there is a trade-off between maintaining image quality and reducing crosstalk due to the limited number of image planes. Physically, it is impossible to use a single phase distribution to create completely independent holographic images for three different wavelengths. This limitation also accounts for the residual chromatic aberration observed in Fig. 3(e) - (h).

The twin image problem in holography is troublesome, especially in the design of binary PDOEs. When the desired diffraction pattern is recorded, its twin image, or the backward wave diffraction pattern, is also recorded. This twin image can sometimes appear in the forward wave diffraction pattern, affecting the upright image in the form of a centrally symmetric pattern. According to the angle spectrum theory [34], the process of holographic reconstruction is

$$U_z(f_x, f_y) = U_0(f_x, f_y) \exp[i\phi_H(f_x, f_y, z)], \tag{9}$$

where $U_0(f_x, f_y)$ is the Fourier transform of the input $u_0(x, y)$, $U_z(f_x, f_y)$ is the Fourier transform of the diffracted field $u_z(x, y)$, and $\phi_H(f_x, f_y, z) = \frac{2\pi z}{\lambda}\sqrt{1 - (\lambda f_x)^2 - (\lambda f_y)^2}$. Due to symmetry, the angular spectrum propagation of the twin image can be written as

$$U_z^*(-f_x, -f_y) = U_0^*(-f_x, -f_y) \exp[-i\phi_H(f_x, f_y, 2z)] \exp[i\phi_H(f_x, f_y, z)]$$
$$= \mathcal{F}\left\{u_0^*(-x, -y) \otimes \left\{\frac{2z}{i\lambda(x^2 + y^2 + 4z^2)} \exp[i\phi_h(f_x, f_y, 2z)]\right\}\right\} \exp[i\phi_H(f_x, f_y, z)]$$
$$= \mathcal{F}\{u_0'(x, y, z)\} \exp[i\phi_H(f_x, f_y, z)], \quad (10)$$

where $\mathcal{F}\{\cdot\}$ denotes Fourier transform and $\otimes$ denotes convolution. The phase related to impulse response is expressed as $\phi_h(f_x, f_y, z) = \frac{2\pi}{\lambda}\sqrt{x^2 + y^2 + z^2}$.

The distribution of input is a decisive factor in determining the shape of the diffraction pattern, rather than the phase delay or complex amplitude modulation of individual pixels. Thus, from our perspective, the similarity between the distributions of $u_0(x, y)$ and $u_0'(x, y, z)$ determines whether the twin image, or $U_z^*(-f_x, -f_y)$, is pronounced and to what extent. If $u_0(x, y)$ and $u_0'(x, y, z)$ are highly similar, the twin image becomes more prominent. Generally, this situation is rare because the degrees of freedom of $u_0(x, y)$ tend towards infinity, making it unlikely to have such strong symmetry. This explains why the twin image is not prominently observed in typical phase retrieval problems. Now, consider a simple and commonly used scenario: a far-field designed monochromatic binary DOE with only 0 or π phase delay for the design wavelength. This configuration has strong symmetry, and the entire DOE can be approximated as a real distribution. Its Fresnel diffraction field $F(x, y)$ exhibits a certain degree of Hermitian symmetry:

$$F(x, y) \approx F^*(-x, -y). \quad (11)$$

So, the twin image, which is identical to the upright image but symmetrical with respect to the coordinate origin, becomes fully visible (see Supplementary Material 1). The SFTM is suitable for diffraction calculations in both near-field and far-field regions, establishing stable constraints in the Fresnel and even Fraunhofer regions. Unlike the Fraunhofer model, which uses a single FFT and possesses an infinitely long DOF, this strong constraint weakens the Hermitian symmetry. Additionally, our strategy, particularly the incorporation of random vectors, further reduces the DOFs. As a result, regardless of the location of the diffraction plane, the distribution of $u_0(x, y)$ remains under strong constraints with a relatively low DOF. Moreover, an achromatic binary PDOE does not have 0 or π distributions for all three wavelengths, further breaking the Hermitian symmetry. Consequently, the final full-color patterns of the PDOEs visually show no appearance of twin images.

However, there is currently no comprehensive mathematical description of the generation, distribution, and impact process of twin images. Under certain conditions, they can be qualitatively decoupled from the upright image, but cannot be fully suppressed in all cases[48,49]. More efforts are needed to address this issue in future studies.

Over all, our method introduces a new paradigm for full-color holography, with significant potential for applications in high information capacity, advanced encryption, enhanced visual effects, and perception. Beyond high-frame-rate full-color holographic videos, our approach is also ideal for polarization-multiplexed metasurfaces, tripling the information capacity of a single polarization channel and maximizing channel multiplexing.

## Conclusion

In summary, we proposed a design pipeline for achromatic single-layer hologram capable of forming specific color images on designated planes by suppressing crosstalk. This approach enables color separation in large-depth 3D CGH or the generation of a scalable full-color image on a specific plane using only a single hologram. We demonstrated full-color holographic video playback at a full 60 Hz refresh rate, tripling the frame rate of current systems. Twin-image-free, achromatic binary security DOEs were successfully designed and fabricated, providing excellent visual effects. Our strategy can be applied to various scenarios, such as product authentication, dynamic color holographic displays, ultra-high-reusability metasurfaces, achromatic metalens and various diffractive optical element designs.

## Supplementary Information
Supplementary Material 1.
Video 1.
Video 2.

## Acknowledgements
The authors would like to thank Yong Tang (Hunan University) for his programming assistance in the production of the holographic video, and Fang Li (Sun Yat-sen University) for providing the cat photos and video source. We are also grateful to Haomiao Zhang (Westlake University) for her valuable suggestions on this paper, and Jing Wang (Hunan University) for his help in getting the experimental results.

## Funding
This work is supported by the Guangzhou Major R&D Funds (ZF202200125) and the Guangzhou Runxin Information Technology Co., Ltd. (SYSU30000-71010519).

## Code and Data Availability
Data are available from the corresponding author upon reasonable request.

## Declarations
**Competing interests**
The authors declare that they have no conflicts of interest.